\documentclass{article}

\usepackage[nonatbib, preprint]{neurips_2020}
\usepackage[utf8]{inputenc} 
\usepackage[T1]{fontenc}    
\usepackage{hyperref}       
\usepackage{url}            
\usepackage{booktabs}       
\usepackage{amsfonts}       
\usepackage{titlesec}       
\usepackage{listings} 
\usepackage{nicefrac}       
\usepackage{microtype}      
\usepackage{graphicx}       
\usepackage{color, colortbl} 
\usepackage{multirow} 
\usepackage{longtable}
\usepackage[most]{tcolorbox} 
\usepackage{lipsum} 
\usepackage[utf8]{inputenc}
\usepackage{tcolorbox}
\usepackage{amsmath}
\usepackage{tcolorbox}

\usepackage{booktabs} 
\usepackage{tabularx} 
\usepackage{geometry} 
\usepackage{lmodern}

\usepackage[utf8]{inputenc}
\usepackage{enumitem}

\definecolor{headercolor}{gray}{0.85}
\definecolor{maxvalue}{rgb}{0.92,1,0.92} 
  \usepackage{listings}       
  \usepackage{xcolor}         
  \usepackage{algorithm}      
  \usepackage{algpseudocode}  

  \lstdefinelanguage{json}{
      basicstyle=\ttfamily\footnotesize,
      commentstyle=\color{gray},
      numbers=left,
      numberstyle=\tiny\color{gray},
      stepnumber=1,
      numbersep=8pt,
      showstringspaces=false,
      breaklines=true,
      frame=single,
      backgroundcolor=\color{gray!5},
      tabsize=2,
      captionpos=b,
      morestring=[b]",
      stringstyle=\color{blue},
      keywordstyle=\color{purple}\bfseries,
      morekeywords={protocolVersion,name,description,url,provider,domain,skills,id,identity,agent_id,public_key,identity_proof,lineage_support,merkle_proof_generation,dpop_binding},
  }
\titleformat{\paragraph}[runin]{\normalfont\normalsize\bfseries}{}{}{}
\titlespacing{\paragraph}{0pt}{0pt}{0.5em}
\tcbuselibrary{skins}

\title{Context Lineage Assurance for Non-Human Identities in Critical Multi-Agent Systems}

\author{%
  Sumana Malkapuram\\
Netflix\\
  \texttt{sumana@skalifes.com} \\
   \And
  Sameera Gangavarapu \\
  Zoetis \\
   \AND
  Kailashnath Reddy Kavalakuntla\\
 Ravenna \\
   \And
   Ananya Gangavarapu \\
    CuratX.ai
}

\begin{document}

\maketitle

\begin{abstract}
The proliferation of autonomous software agents necessitates rigorous frameworks
for establishing secure and verifiable agent-to-agent (A2A) interactions,
particularly when such agents are instantiated as non-human identities (NHIs).
We extend the A2A paradigm \cite{a2a,block-a2a} by introducing a cryptographically grounded mechanism
for lineage verification, wherein the provenance and evolution of NHIs are
anchored in append-only Merkle tree structures modeled after Certificate
Transparency (CT) logs. Unlike traditional A2A models that primarily secure
point-to-point interactions, our approach enables both agents and external
verifiers to cryptographically validate multi-hop provenance, thereby ensuring
the integrity of the entire call chain. A federated proof server acts as an
auditor across one or more Merkle logs, aggregating inclusion proofs and
consistency checks into compact, signed attestations that external parties can verify without access to the full execution trace. In parallel, we augment the A2A
agent card to incorporate explicit identity verification primitives, enabling
both peer agents and human approvers to authenticate the legitimacy of NHI
representations in a standardized manner. Together, these contributions establish
a cohesive model that integrates identity attestation, lineage verification, and
independent proof auditing, thereby advancing the security posture of inter-agent
ecosystems and providing a foundation for robust governance of NHIs in regulated
environments such as FedRAMP.
\end{abstract}

\section{Introduction}

The accelerating deployment of autonomous agents across diverse computational domains, ranging from financial services and healthcare to supply chain logistics and cyber-defense, has prompted an urgent re-examination of the security assumptions underpinning agent-to-agent (A2A) communication. These agents, instantiated as non-human identities (NHIs), increasingly assume responsibilities once reserved for human actors, including negotiation, decision-making, and the execution of binding commitments. As their operational scope expands, so too does the attack surface, particularly in the absence of robust identity and lineage verification mechanisms.

Existing A2A frameworks \cite{a2a} primarily emphasize message integrity, confidentiality, and authentication at the transport layer. While such measures protect against conventional threats (e.g., eavesdropping or message tampering), they fall short in addressing deeper questions of provenance and legitimacy. Specifically, current A2A specifications lack the ability to:
\begin{enumerate}
    \item \textbf{Verify lineage:} Agents cannot reliably establish the historical authenticity of their counterpart’s identity or confirm that the presented identity has not been revoked, superseded, or tampered with across its lifecycle. Without lineage verification, adversaries can mount persistent impersonation or replay attacks that are indistinguishable from legitimate interactions.
    \item \textbf{Authenticate NHIs beyond credentials:} Most frameworks rely on cryptographic keys or digital certificates to prove identity, but such primitives are brittle in the absence of contextual metadata and verifiable attestation. An NHI’s ``agent card'' may contain descriptive fields, yet these are not cryptographically bound to verifiable claims, leaving them susceptible to forgery or omission.
\end{enumerate}

This gap between current A2A practice and the security requirements of real-world agent ecosystems motivates the present work. We contend that identity verification and lineage integrity must be treated as first-class properties of agent-to-agent interactions, rather than as peripheral concerns. Our proposed enhancements, rooted in the use of Merkle trees for lineage verification and explicit identity verification in agent cards, seek to bridge this gap by embedding verifiability and resilience directly into the A2A substrate. In doing so, we aim to establish a foundation for the next generation of secure, auditable, and interoperable agent ecosystems, in which NHIs can be trusted not only for the correctness of their immediate actions, but also for the integrity of their histories.

\section{Background}

\subsection{Agent-to-Agent (A2A) Communication}
Agent-to-Agent (A2A) communication protocols are designed to enable autonomous entities to exchange information, negotiate tasks, and coordinate actions without direct human intervention. Existing frameworks typically focus on transport-layer security, ensuring message confidentiality, integrity, and endpoint authentication through mechanisms such as TLS or public key infrastructures (PKI). While effective for securing communication channels, these protocols often neglect higher-order requirements, such as establishing provenance of identities and verifying the historical continuity of credentials. As autonomous agents expand their operational roles into sensitive domains, these omissions have become increasingly untenable.

\subsection{Non-Human Identities (NHIs)}
The concept of non-human identities (NHIs) formalizes the representation of autonomous agents as distinct, verifiable entities within digital ecosystems. NHIs may correspond to software agents, machine learning models, or cyber-physical systems acting on behalf of an organization. Unlike human users, NHIs lack conventional identity attributes (e.g., biometrics) and instead rely on cryptographic credentials and metadata to assert legitimacy. This distinction introduces new challenges: NHIs must not only authenticate themselves but also provide verifiable guarantees of continuity, revocation, and trust delegation. Current identity systems, which are designed primarily for human principals, do not adequately address these requirements.

\subsection{Merkle Trees and Certificate Transparency (CT) Logs}
Merkle trees are cryptographic data structures that allow efficient, tamper-evident verification of large datasets by committing to the dataset’s state in a single hash root. They have been widely adopted in log-based transparency systems. Certificate Transparency (CT) logs \cite{laurie2014certificate}, in particular, employ Merkle trees to provide append-only, auditable records of issued TLS certificates. These logs enable clients to detect mis-issuance and unauthorized modifications, establishing accountability in public key infrastructures. By analogy, applying Merkle-tree based lineage logs to NHIs enables the recording and verification of identity creation, updates, and revocations in a manner that is both efficient and tamper-resistant.

\subsection{Agent Cards and Identity Verification}
The notion of an ``agent card'' has emerged as a convenient abstraction for describing the identity and capabilities of autonomous agents. An agent card typically contains metadata such as the agent’s name, purpose, and cryptographic credentials. However, current implementations treat this card largely as a descriptive artifact, without strong cryptographic binding between its contents and verifiable proofs. As a result, malicious actors can forge, omit, or alter metadata without detection. Strengthening the agent card with cryptographically verifiable identity attributes, attested by trusted authorities, would elevate it from a descriptive profile to a robust identity verification mechanism.

\section{Security Goals and Design Principles}
The security of agent-to-agent (A2A) ecosystems rests not only on the confidentiality and integrity of messages, but also on the authenticity, lineage, and verifiability of the non-human identities (NHIs) involved. Based on the shortcomings identified in current A2A frameworks and the adversarial scenarios outlined in our threat model, we define the following security goals and design principles for the proposed system.
\subsection{Lineage Verifiability}
\textbf{Goal:} Agents must be able to independently verify the historical continuity of an identity and action lineage. \\
\textbf{Rationale:} Without action lineage, the downstream agents miss the necessary information to take a decision. \\
\textbf{Principle:} Employ cryptographic append-only structures (Merkle trees) to record all NHI lifecycle events, ensuring tamper-evidence and efficient validation of action events.

\subsection{Strong Identity Attestation}
\textbf{Goal:} NHIs must present identity attributes that are cryptographically bound to verifiable proofs, not merely descriptive metadata. \\
\textbf{Rationale:} Current agent cards provide weak assurances, allowing forgery, omission, or selective disclosure attacks. \\
\textbf{Principle:} Extend the agent card with mandatory identity proofs (e.g., cryptographic signatures, verifiable credentials, and lineage commitments) to make claims inseparable from their attestations.

\subsection{Minimal Disclosure and Privacy Preservation}
\textbf{Goal:} Agents must be able to verify the integrity of claims made by other agents without requiring the disclosure of underlying sensitive data. \\
\textbf{Rationale:} The attributes, history, and metadata associated with an NHI can reveal sensitive strategic, operational, or proprietary information. Verification protocols must be designed to prevent this data leakage. \\
\textbf{Principle:} Combine Merkle trees with \textbf{Zero-Knowledge Proofs (ZKPs)} to create privacy-preserving attestations.
\begin{itemize}
    \item \textbf{Commitment:} An agent's sensitive attributes are cryptographically committed to and placed as leaves in its Merkle tree.
    \item \textbf{Proof Generation:} To prove a claim, the agent generates a \textit{ZKP} that proves a property about a committed attribute (\textit{e.g., "proves the value of a credit line is \> \$1M"}) and a \textit{Merkle proof} that proves this commitment is part of its official history.
    \item \textbf{Verification:} The verifier checks these two proofs. They are convinced the claim is true without ever seeing the actual sensitive data (\textit{the exact credit line value}). This creates a truly selective and private disclosure mechanism.
\end{itemize}

\subsection{Scalability and Efficiency}
\textbf{Goal:} Keep the cryptographic verification path lightweight and efficient, ensuring that proof checks do not become a bottleneck at scale. \\
\textbf{Rationale:} The integrity of the call chain depends on proving inclusion and consistency of events in transparency logs. If these proofs were expensive, the system would not be practical for high-throughput A2A interactions. Other runtime costs such as DID resolution, network latency, and state management are important but fall outside the scope of this complexity analysis. \\
\textbf{Principle:} Optimize for logarithmic verification complexity in the cryptographic path. Merkle inclusion and consistency proofs scale as $O(\log n)$ in the number of log entries, while signature verifications remain constant time. Proof bundles should be compact so that cryptographic verification can scale to millions of events. Non-cryptographic overheads are treated as deployment concerns and are out of scope for this design principle.
\section{Architectural Overview}

Our architecture strengthens agent-to-agent (A2A) communication by embedding verifiable identity and lineage guarantees into the core protocol. It is organized around two key elements that together enable secure, auditable, and scalable interactions among non-human identities (NHIs):

\begin{enumerate}
    \item \textbf{Cryptographically verifiable identity of NHIs.} 
    The architecture extends the standard A2A Agent Card with cryptographic bindings that tie each agent’s identifier to its public key, signed proofs, and declared capabilities. This ensures that every NHI can be uniquely and verifiably represented across distributed ecosystems.
    
    \item \textbf{System mechanics for action lineage.} 
    The architecture incorporates a tamper-evident mechanism for recording and validating agent actions at scale. By committing each event to an append-only Merkle-based Lineage Store and enabling proof generation via a Proof Server, the system provides efficient $O(\log n)$ verification of call chains in multi-agent environments.
\end{enumerate}

Together, these elements establish an architecture where NHIs are not only identifiable with strong cryptographic evidence but also accountable for the sequence of actions they perform. The identity extension provides the root of trust, while the lineage mechanism builds on it to deliver scalable provenance across complex agent ecosystems.

\subsection{Cryptographically Verifiable Identity of NHIs}
\label{sec:identity}

Our framework extends the base agent-to-agent (A2A) protocol with identity verification capabilities that bind each non-human identity (NHI) to cryptographic proofs. This ensures that every agent representation is both unique and verifiable, forming the foundation for constructing tamper-evident call chains.

\textbf{A2A Protocol Identity Extension.} 
The extension augments the standard A2A Agent Card with an additional \texttt{identity} block containing cryptographically bound identifiers and proofs. This design preserves backwards compatibility with existing A2A implementations while enabling verifiable identity attestation.

\textbf{Enhanced Agent Card Structure.} 
The enhanced Agent Card introduces mandatory identity fields that bind an NHI’s key material to verifiable claims. An example generic agent card is shown in Listing~\ref{lst:agentcard}.

\lstset{language=json, caption={Enhanced A2A Agent Card with Identity Extension}, label={lst:agentcard}}
\begin{minipage}{\linewidth}
\begin{lstlisting}
{
  "protocolVersion": "1.0",
  "name": "workflow-approver",
  "description": "Automated approval and routing agent",
  "url": "https://agents.example.com/workflow-approver",
  "provider": {
    "name": "ExampleOrg",
    "domain": "example.com"
  },
  "skills": [
    {
      "id": "approve-task",
      "name": "Task Approval",
      "description": "Approves and routes workflow tasks"
    }
  ],
  "identity": {
    "agent_id": "aid://sha256(AB12CD34||example.com||1640995200)",
    "public_key": "ed25519:302a300506032b6570032100AB12CD34...",
    "identity_proof": "ed25519:4F8A7B2C9D1E3F5A6B8C9E0F2A4D6E8B...",
    "lineage_support": {
      "merkle_proof_generation": true,
      "dpop_binding": true
    }
  }
}
\end{lstlisting}
\end{minipage}

\textbf{Identity Field Specifications.} 
\begin{itemize}
    \item \textbf{\textit{agent\_id}} is derived as 
    \[
    \textit{SHA-256}(\textit{public\_key} \,\|\, \textit{provider.domain} \,\|\, \textit{timestamp}),
    \]
    ensuring global uniqueness and preventing impersonation.
    \item \textbf{\textit{public\_key}} stores the agent’s \textit{Ed25519} public key, which enables external parties to validate signatures and cryptographic bindings.
    \item \textbf{\textit{identity\_proof}} contains
    \[
    \textit{Sign}_{priv}(agent\_id \,\|\, skills),
    \]
    demonstrating key possession and binding the identity to its declared capabilities.
    \item \textbf{\textit{lineage\_support}} specifies whether the agent can generate Merkle proofs \cite{merkle1987} and bind to DPoP\cite{dpop} tokens for replay attack mitigation.
\end{itemize}

\textbf{Registration and Verification.} 
During initialization, an agent generates an \textit{Ed25519} keypair, computes its \texttt{agent\_id}, signs the identity proof, and registers the enhanced Agent Card with a discovery service. Other agents retrieve the card from the provider’s \texttt{/.well-known/agent-card.json}, recompute the hash, and validate the proof. This process ensures that every NHI identity in the system is anchored in cryptographic evidence, creating the basis for the action lineage model introduced in the next subsection.

\subsection{Recording and Verifying Action Lineage at Scale}

While cryptographic identity provides the foundation of trust, identity alone is not sufficient for accountability. In complex multi-agent environments, downstream agents must be able to verify not only \emph{who} performed an action, but also \emph{what} was done, \emph{when} it occurred, and under which governing context. To address this requirement, our architecture incorporates a tamper-evident action lineage system that records and validates every agent decision in a scalable and efficient manner.

\textbf{Action Lineage.} 
Action lineage represents a cryptographically committed history of agent activity, capturing the essential metadata required for verification without exposing sensitive payloads. Each recorded event contains just enough context to validate authenticity, continuity, and configuration state, thereby minimizing disclosure while ensuring verifiability.\\
\textbf{Core Components.} 
The lineage subsystem consists of two tightly coupled services:
\begin{enumerate}
    \item \textbf{Lineage Store (LS).} An append-only, authenticated log that maintains lifecycle and action events for NHIs. The LS is optimized for high-throughput \texttt{Append} operations and read-optimized \texttt{Prove} operations, with data stored as Merkle trees to ensure tamper-evidence.

    \begin{itemize}
        \item \emph{Structure.} A dynamic, append-only Merkle tree is maintained over event leaves. Each leaf is computed as
        \[
        L_i = H(0x00 \,\|\, c_{e_i}),
        \]
        where $c_{e_i}$ is the canonical encoding of event $e_i$. Internal nodes are computed as
        \[
        N = H(0x01 \,\|\, L \,\|\, R),
        \]
        with domain separation bytes ($0x00$ for leaves, $0x01$ for internal nodes) preventing cross-type collisions.
        
        \item \emph{Snapshot (checkpoint).} After $n$ appends, the LS issues a \emph{Signed Tree Head (STH)}:
        \[
        STH_n = \text{Sign}_{LS}(n, root_n, wallclock\_t, monotonic\_ctr, log\_id),
        \]
        where $root_n$ is the Merkle root after $n$ leaves, $wallclock\_t$ provides temporal anchoring, $monotonic\_ctr$ enforces ordering, and $log\_id$ identifies the lineage log instance.
    \end{itemize}

    \item \textbf{Proof Server (PS).} A stateless (or cache-heavy) service responsible for returning proofs of correctness. It provides:
    \begin{itemize}
        \item \emph{Inclusion proofs:} minimal audit paths verifying that a given event is included in a snapshot $(i, n)$.
        \item \emph{Consistency proofs:} evidence that one snapshot $n_1$ is a prefix of another $n_2$, without revealing raw leaves.
        \item \emph{Multiproofs:} efficient, de-duplicated node sets sufficient to validate a set $S = \{i_k\}$ within snapshot $n$, with size approximately 
        \[
        O\big(\log n + k \log \tfrac{n}{k}\big).
        \]
        \item \emph{Batching:} query coalescing by $(n, \text{subtree})$ and serving from a hot cache of recent Signed Tree Heads (STHs).
    \end{itemize}
\end{enumerate}

\textbf{Event Model and Commitments.} 
Each agent commits its activity to the lineage store in the form of Merkle leaves. The \texttt{agent\_id} used here is the same cryptographically derived identifier introduced in Section~\ref{sec:identity}, ensuring that every action is linked back to a verifiable NHI identity. The canonical event model is defined as:

\begin{lstlisting}[language=json, caption={Canonical Event Model}, label={lst:eventmodel}]
{
  "agent_id": "aid://sha256(pubkey||domain||ts)",      
  "action_id": "uuid-1234",
  "ts": 1710525600,
  "action_type": "approve_invoice",
  "context_hash": "H(policy/version/etc.)",
  "agent_sig": "signature(agent_id||action_id||...)"
}
\end{lstlisting}

From this model, the leaf hash is computed as:
\[
\text{leaf\_hash} = H(\text{encode(event)}).
\]

The \texttt{agent\_sig} ensures that each committed event is signed with the agent’s private key, binding the event to the cryptographically verifiable identity defined in the Agent Card. This creates an unbroken link between identity and action lineage.


\section{Framework: Lineage Registration and Verification}

The second element of our architecture builds directly on the identity foundation established in Section~\ref{sec:identity}. While the enhanced Agent Card provides each non-human identity (NHI) with a cryptographically verifiable identifier, the lineage subsystem ensures that every action attributed to that identity is anchored in an immutable log. This tight coupling between \texttt{agent\_id} and the lineage store allows downstream agents to verify both the authenticity of the actor and the integrity of its recorded behavior.

\subsection{Lineage Registration.} 
When an agent executes an action, it generates an event in the canonical format described in Listing~\ref{lst:eventmodel}. The event is signed with the agent’s private key, binding the \texttt{agent\_id} (from the Agent Card) to the specific action, timestamp, and context. The resulting leaf hash is appended to the Lineage Store (LS), which incorporates the event into its append-only Merkle tree. The LS periodically issues a Signed Tree Head ($STH_n$), attesting to the state of the log after $n$ appends. This ensures that every action is durably committed, tamper-evident, and provable against the current Merkle root.

\subsection{Lineage Verification and Provenance.} 
During invocation, a downstream agent must establish whether the upstream caller is both authentic and part of the legitimate call chain. This process involves three verification steps:

\begin{enumerate}
    \item \emph{Merkle Proofs.} The verifier queries the Proof Server (PS) to obtain inclusion proofs for one or more upstream actions. The PS returns audit paths that allow the verifier to check, against the latest $STH_n$, whether the claimed events are indeed present in the lineage store.
    
    \item \emph{Identity Binding.} For each verified event, the verifier retrieves the upstream agent’s Agent Card (\texttt{/.well-known/agent-card.json}) and recomputes the \texttt{agent\_id}. The verifier then validates the \texttt{agent\_sig} in the event against the public key in the card. This confirms that the upstream action was signed by the same cryptographically bound identity introduced in Section~\ref{sec:identity}.
    
    \item \emph{Call Chain Provenance.} By iterating this process across all upstream nodes, the verifier reconstructs the call chain and ensures that each participant is both recorded in the log and authenticated via its Agent Card. Any inconsistency—such as a missing proof, a mismatched signature, or an invalid identifier—causes the verifier to terminate the interaction and refuse to append new events.
\end{enumerate}

This framework ensures that the verifiable identity of NHIs (the first architectural element) and their tamper-evident action history (the second architectural element) are inseparably linked. Together, they provide a foundation for secure, auditable, and resilient multi-agent ecosystems, where every action can be traced back to a provable origin.

\subsection{Prelude: Motivation for Human-in-the-Loop Lineage in FedRAMP}

The presented use-case emphasizes a realistic scenario in which a cloud service
provider seeks to achieve FedRAMP authorization by combining the efficiency of
autonomous software agents with the accountability of human decision makers.
FedRAMP is not merely a technical certification, but a regulatory framework
rooted in risk acceptance and trust. While automated agents can accelerate the
collection of evidence, execution of vulnerability scans, and assembly of the
System Security Plan (SSP), the framework explicitly requires that human
officials---such as Authorizing Officials (AOs), compliance leads, and
independent Third-Party Assessment Organizations (3PAOs)---review, validate,
and formally approve key milestones.

The core emphasis of this use-case is on ensuring that \emph{every action and
every approval} is recorded as an immutable event in a cryptographic lineage
log. This guarantees that:

\begin{itemize}
  \item Autonomous agents cannot fabricate or modify evidence without detection.
  \item Human approvals, such as risk acceptance or SSP validation, are bound to
        the evidence they pertain to and cannot later be repudiated.
  \item Independent assessors (3PAOs) can replay the call chain to confirm that
        the evidence provided in the SSP was anchored at the time of collection
        and endorsed by the proper authorities.
\end{itemize}

In short, the real-world requirement addressed here is the ability to provide
\textbf{cryptographically verifiable chain-of-custody across both automated and
human actors}. This satisfies FedRAMP’s emphasis on accountability, non-
repudiation, and independent auditability, while still leveraging automation to
reduce the operational burden of assembling and maintaining the authorization
package.

We model the FedRAMP certification workflow as a sequence of cryptographically
anchored events, produced by both autonomous agents (non-human identities, NHIs)
and human trust anchors. Each event is committed to an append-only lineage log
implemented as a Merkle tree, with periodic signed tree heads (STH) for external
verifiability.

\subsection{Notation}

\begin{itemize}
  \item $E_i$ : Event at step $i$.
  \item $L_i$ : Leaf hash at step $i$.
  \item $H(\cdot)$ : Cryptographic hash (e.g., SHA-256).
  \item $\text{sig}_X(m)$ : Signature over message $m$ by actor $X$.
  \item $PK_X$ : Public key of actor $X$ (agent or human).
  \item $R_t$ : Root hash of lineage log after $t$th event append.
  \item $\pi(E_i, R_t)$ : Merkle inclusion proof showing $E_i$ is in tree with root $R_t$.
\end{itemize}

\subsection{Workflow Events}
\label{sec:workflow}

\paragraph{Step 0: Human AO boundary approval}
The Authorizing Official (AO) or system owner formally defines the system boundary
and selects the applicable FedRAMP baseline (Low, Moderate, or High).
This approval anchors the workflow in human accountability, serving as the root
of trust for all subsequent actions. Without this signed approval, no automated
agent may initiate activity. This step satisfies FedRAMP controls PL-2
(System Security Plan) and CA-6 (Authorization), which require explicit AO
approval of scope and accountability.

\[
E_0 = \big(\textsf{boundary\_approval}, ts_0, \text{sig}_{AO}(\cdot)\big)
\]
\[
R_0 = H(E_0)
\]

\paragraph{Step 1: A1 readiness start}
The Readiness Coordinator agent initiates the FedRAMP workflow by generating a
\texttt{readiness\_start} event that references the AO's signed boundary approval.
This establishes the first agent-driven lineage entry and cryptographically binds
the automation sequence to the human decision. The event demonstrates that the
automation only proceeds under authorized conditions, supporting AU-2 (Audit Events)
and AU-10 (Non-repudiation).

\[
E_1 = \big(\textsf{readiness\_start}, prev=E_0, ts_1, \text{sig}_{A1}(\cdot)\big)
\]
\[
R_1 = H(R_0 \,\|\, E_1)
\]

\textit{Commentary — Readiness Coordinator Agent (A1~\ref{lst:A1}).}
The Readiness Coordinator agent acts as the entry point into the automated FedRAMP workflow. Its role is to translate the human Authorizing Official’s boundary approval into a machine-verifiable event, ensuring that automation only proceeds when explicitly authorized. This agent has minimal scope: it cannot fabricate evidence or bypass controls, but simply records the start of a compliance workflow and binds it cryptographically to the AO’s approval. By doing so, A1 enforces the principle that no subsequent automated step is valid unless rooted in a legitimate human decision.

\paragraph{Step 2: A2 evidence collection}

The Evidence Harvester agent queries system-of-record APIs (e.g., AWS Config,
Azure Resource Graph, GCP Asset Inventory) to collect asset inventories,
IAM policies, and configuration baselines. The resulting artifacts are hashed
and anchored into the lineage log. This guarantees that evidence originates
from authoritative sources rather than manual or fabricated inputs. This step
addresses controls CM-8 (Component Inventory) and AU-6 (Audit Review).

\[
E_2 = \big(\textsf{collect\_inventory}, prev=E_1, ts_2, H(\text{artifacts}), \text{sig}_{A2}(\cdot)\big)
\]
\[
R_2 = H(R_1 \,\|\, E_2)
\]

\textit{Commentary — Evidence Harvester Agent (A2~\ref{lst:A2}).}
Evidence Harvester agent is responsible for interfacing directly with
system-of-record (SOR) APIs across cloud environments to obtain authoritative data
such as asset inventories, IAM policies, and configuration baselines. Its
scope is deliberately narrow: it cannot generate or alter evidence, only
collect and hash what is published by trusted infrastructure sources. By
anchoring these artifacts into the lineage log, A2 ensures that downstream
analysis, scanning, and reporting are rooted in verifiable ground truth
rather than manually curated or potentially fabricated inputs. This agent
effectively bridges automated compliance workflows with the integrity of the
underlying infrastructure state.

\paragraph{Step 2a: Human compliance approval:}

A compliance officer reviews the harvested evidence and confirms that the scope matches the previously approved system boundary. This step prevents "scope drift" where in-scope resources might be omitted from the package, out-of-scope resources might be inadvertently included, or boundary definitions might have changed since initial approval. The approval is digitally signed and appended to the lineage log, satisfying PL-2 (System Security Plan) and CA-2 (Security Assessments).

\[
E_{2a} = \big(\textsf{inventory\_approval}, prev=E_2, ts_{2a}, \text{sig}_{CL}(\cdot)\big)
\]
\[
R_{2a} = H(R_2 \,\|\, E_{2a})
\]

\paragraph{Step 3: A3 scanning}
Scanner Orchestrator agent executes vulnerability and configuration scans
against the inventory defined in Step 2. Results are hashed and committed to
the lineage log, providing immutable proof that scans were executed against
the actual boundary. This satisfies RA-5 (Vulnerability Scanning) and SI-2
(Flaw Remediation).

\[
E_3 = \big(\textsf{scan\_results}, prev=E_{2a}, ts_3, H(\text{report}), \text{sig}_{A3}(\cdot)\big)
\]
\[
R_3 = H(R_{2a} \,\|\, E_3)
\]

\textit{Commentary — Scanner Orchestrator Agent (A3~\ref{lst:A3}).}
Scanner Orchestrator agent is tasked with executing vulnerability and configuration scans over the inventory collected by A2. Its primary function is to ensure that scans are performed systematically, reproducibly, and against the exact boundary previously approved and recorded. A3 validates scan completeness, handles scan failures and retries, and ensures coverage of all asset types within scope. A3 does not interpret or remediate findings; instead, it produces signed results that are cryptographically hashed and logged, providing immutable evidence that the scans were executed. This separation of duties keeps the agent narrowly focused on orchestration and verifiable execution, leaving risk decisions to human approvers in later steps.

\paragraph{Step 3a: Human security officer approval}:
The Security Officer reviews the scan findings. In cases where vulnerabilities
cannot be remediated immediately, the officer may accept the residual risk or
document compensating controls. This approval is signed and recorded in the
lineage log, providing evidence of human accountability for risk decisions,
as required by CA-6 (Authorization) and PL-2 (System Security Plan).
\[
E_{3a} = \big(\textsf{risk\_acceptance}, prev=E_3, ts_{3a}, \text{sig}_{SO}(\cdot)\big)
\]
\[
R_{3a} = H(R_3 \,\|\, E_{3a})
\]

\paragraph{Step 4: A4 SSP build}
The SSP Packager agent assembles the System Security Plan (SSP), citing
previous evidence events (inventory and scans). The generated SSP sections are
hashed and anchored in the lineage log. This ensures the SSP is verifiably tied
to underlying evidence rather than manually drafted text. This supports PL-2
(System Security Plan) and AU-6 (Audit Review).
\[
E_4 = \big(\textsf{build\_ssp}, cites=\{E_2,E_3\}, prev=E_{3a}, ts_4, H(\text{ssp}), \text{sig}_{A4}(\cdot)\big)
\]
\[
R_4 = H(R_{3a} \,\|\, E_4)
\]

\textit{Commentary — SSP Packager Agent (A4~\ref{lst:A4}).}
The SSP Packager agent is responsible for assembling the System Security Plan
(SSP) by drawing directly on prior evidence events, such as inventory data
from A2 and scan results from A3. Its role is not to author new content but
to ensure that the SSP is mechanically bound to verifiable, logged evidence.
This design prevents drift between documentation and actual system state,
providing auditors with cryptographically anchored assurance that the SSP
reflects reality. A4 therefore acts as a compliance-focused assembler,
transforming raw evidence into structured documentation that can be reviewed
and approved by human compliance officers.

\paragraph{Step 4a: Human compliance lead approval}:
A compliance officer reviews the SSP generated by the agent to confirm accuracy
and completeness. The signed approval is anchored into the lineage log,
satisfying FedRAMP’s requirement that humans validate the correctness of the SSP.
This step enforces accountability and non-repudiation.
\[
E_{4a} = \big(\textsf{ssp\_approval}, prev=E_4, ts_{4a}, \text{sig}_{CL}(\cdot)\big)
\]
\[
R_{4a} = H(R_4 \,\|\, E_{4a})
\]

\paragraph{Step 5: A5 capsule + 3PAO SAR}
The 3PAO Liaison agent packages the cited evidence (inventory, scans, SSP)
along with Merkle inclusion proofs into an ``evidence capsule'' for delivery
to the independent assessor. The event is logged to establish provenance of
what was provided. This step prepares the authorization package for independent
review.
\[
E_5 = \big(\textsf{publish\_capsule}, cites=\{E_2,E_3,E_4\}, prev=E_{4a}, ts_5, \text{sig}_{A5}(\cdot)\big)
\]
\[
R_5 = H(R_{4a} \,\|\, E_5)
\]

\textit{Commentary — 3PAO Liaison Agent (A5~\ref{lst:A5}).}
The 3PAO Liaison agent packages the collected evidence such as inventories,
scan results, and the generated SSP together with their Merkle inclusion
proofs into an auditable “evidence capsule.” Its purpose is to ensure that
external assessors receive a complete, tamper-evident package that can be
independently verified against the transparency log. A5 does not alter or
reinterpret the evidence. It functions purely as a trusted courier and
provides verifiable provenance of what was submitted for external review. By
formalizing the handoff to independent assessors, this agent establishes a
clear boundary of accountability between internal automation and external
audit.

\paragraph{Step 5a: 3PAO Human SAR Approval}
Third-Party Assessment Organization (3PAO) assessors independently verify the
capsule against the transparency log, confirm inclusion proofs, and generate a
Security Assessment Report (SAR). The signed SAR is anchored into the lineage log,
satisfying CA-2 (Security Assessments) and AU-10 (Non-repudiation).
\[
E_{5a} = \big(\textsf{sar\_signed}, prev=E_5, ts_{5a}, \text{sig}_{3PAO}(\cdot)\big)
\]
\[
R_{5a} = H(R_5 \,\|\, E_{5a})
\]

\paragraph{Step 6: Agency AO ATO decision}
The Agency AO or Joint Authorization Board (JAB) officials review the SAR and
SSP, weighing mission needs against residual risks. They issue a signed
Authorization to Operate (ATO) decision, which is anchored in the lineage log.
This final human-in-the-loop checkpoint fulfills CA-6 (Authorization) and
provides the ultimate trust anchor in the FedRAMP lifecycle.
\[
E_6 = \big(\textsf{ato\_decision}, prev=E_{5a}, ts_6, \text{sig}_{AO}(\cdot)\big)
\]
\[
R_6 = H(R_{5a} \,\|\, E_6)
\]

\begin{figure}[!ht]
  \centering
  \includegraphics[width=\linewidth]{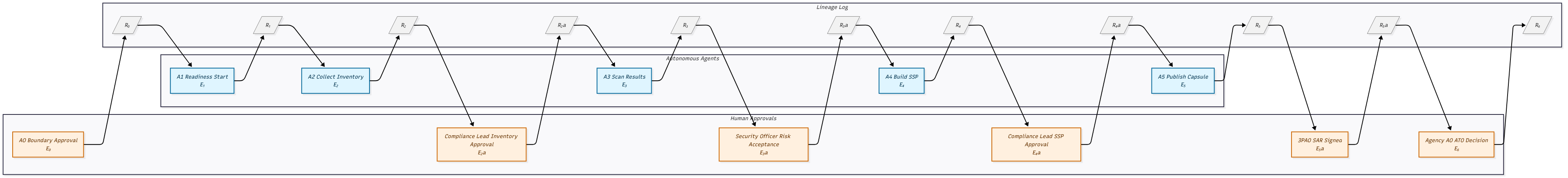}
  \caption{Lineage of workflow events $E_0 \rightarrow E_6$, showing agent
  actions, human approvals, and log roots $R_i$.}
  \label{fig:workflow}
\end{figure}

Each agent described above is accompanied by a structured Agent Card, included in Appendix~\ref{appendix:agentcards}, which specifies its identity, skills, and lineage support features.
---
\subsection{Call Chain Verification}

We now formalize the verification process that ensures the integrity of the
entire call chain. For any cited event $E_i$, an external verifier (e.g., a
3PAO or Agency AO) must establish that (a) the event was created by the
claimed actor, (b) the event is immutably anchored in one or more Merkle logs,
and (c) the anchoring has been validated by the Proof Server acting as a
federated auditor. By chaining these conditions across all events
$E_0 \rightarrow E_1 \rightarrow \cdots \rightarrow E_n$, the verifier can
reconstruct a tamper-evident lineage of both autonomous agent actions and
human approvals.

\paragraph{Step 1: Actor Authenticity.}
Each event $E_i$ is signed by the actor that produced it, yielding
$\text{sig}_{actor}(E_i)$. The verifier retrieves the actor’s public key
$PK_{actor}$ from its registered \emph{Agent Card} (for autonomous agents) or
from an \emph{identity registry} (for human approvers). The verifier then
checks:
\[
\text{Verify}\big(\text{sig}_{actor}(E_i), PK_{actor}\big) = \top
\]
ensuring that $E_i$ was produced by the claimed entity and has not been
modified in transit. The binding between $PK_{actor}$ and the actor’s
identity is assumed to be established through prior enrollment and is
auditable by external verifiers. This step provides authenticity and
non-repudiation at the granularity of individual events.

\paragraph{Step 2: Proof Server Attestation.}
The verifier obtains an \emph{audited proof package} $\mathcal{P}(E_i)$ from
the Proof Server:
\[
\mathcal{P}(E_i) = \big(E_i, L_i, \{STH_t^k\}, \{\pi(E_i, R_t^k)\},
\text{sig}_{PS}(\cdot)\big)
\]
where $L_i = H(E_i)$ is the leaf hash, $STH_t^k$ are signed tree heads from
underlying Merkle logs, and $\pi(E_i, R_t^k)$ are the inclusion proofs showing
that $L_i$ is contained in the tree with root $R_t^k$.
The Proof Server validates both inclusion and consistency across logs before
signing the package.

\paragraph{Step 3: Proof Server Signature.}
The verifier checks:
\[
\text{Verify}\big(\text{sig}_{PS}(\mathcal{P}(E_i)), PK_{PS}\big) = \top
\]
ensuring that the Proof Server has endorsed the correctness of the inclusion
proofs and the validity of the underlying log roots.

\paragraph{Step 4: Workflow Integrity.}
Since each event $E_i$ contains a pointer $prev$ to its predecessor, the
verifier can recursively validate the entire chain:
\[
E_0 \rightarrow E_1 \rightarrow \cdots \rightarrow E_n
\]
by applying Steps 1--3 to each $E_i$ in sequence. Human approvals
($E_0, E_{2a}, E_{3a}, E_{4a}, E_{5a}, E_6$) appear as signed events in this
chain, serving as non-repudiable trust anchors. Autonomous agent events
($E_1, E_2, E_3, E_4, E_5$) provide cryptographically verifiable evidence
collection, scanning, and packaging operations. The combination ensures that
the entire workflow is both \emph{cryptographically bound to its predecessors}
and \emph{independently auditable}.
\begin{figure}[H]
  \centering
  \includegraphics[width=\linewidth]{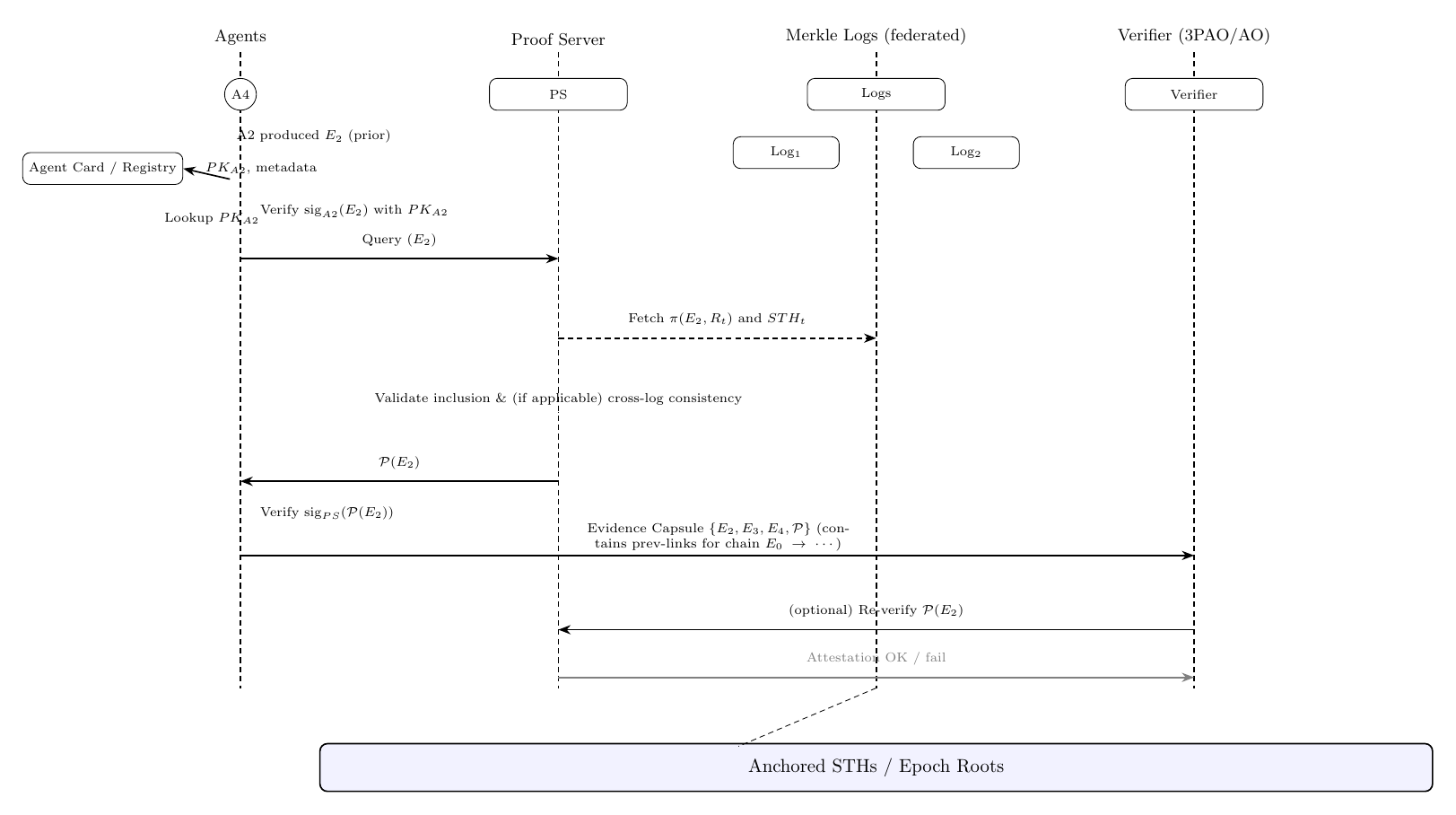}
  \caption{Call Chain Verification for a single event $E_2$: actor authenticity via $PK_{A2}$; audited proof package $\mathcal{P}(E_2)$ from the Proof Server using a federated Merkle Logs service; local verification of $\text{sig}_{PS}$; and capsule delivery to the 3PAO with optional re-verification.}
  \label{fig:callchain-verify}
\end{figure}

By enforcing these verification steps, an external assessor can reconstruct a
tamper-evident lineage graph from boundary definition through ATO decision. This
establishes accountability for both human and autonomous actors and provides
the cryptographic assurance required for regulated environments such as FedRAMP.

\section*{Future Considerations}

One potential optimization, which we consider out of scope for this paper, is
\emph{checkpointed compaction of lineage logs}. This technique could help reduce
historical proof sizes and lower verification costs in high-throughput,
multi-agent workflows, while preserving the append-only properties of the
primary log.
\paragraph{Motivation.}
Even with minimal per-event payloads, high-throughput, multi-hop A2A workflows produce long lineage chains. To reduce historical proof sizes and serving costs \emph{without mutating history}, we sketch a non-destructive compaction scheme.

\paragraph{Approach.}
A secondary, \emph{checkpointed} Merkle tree \(C\) is maintained by the Proof Server (PS), trailing the primary append-only log \(T\) by a safety lag.
At a fixed cadence (time- or volume-based), the PS partitions \(T\) into closed segments \(S_e = [i_e, j_e]\) and commits a \emph{segment root}
\[
R_e \;=\; \mathrm{MerkleRoot}\big(\mathrm{Leaves}(i_e..j_e)\big).
\]
The sequence \((R_1,\dots,R_k)\) forms \(C\), whose head \(\mathrm{CTH}_k\) is published with metadata (epoch boundaries, hash identifiers, sizing policy).
The primary Lineage Store remains unchanged and authoritative.

\paragraph{Verification.}
Inclusion reduces to a two-stage proof:
(i) an intra-segment path from a leaf \(\ell \in S_e\) to \(R_e\);
(ii) an inter-segment path from \(R_e\) to \(\mathrm{CTH}_k\).
Optionally, a consistency proof from the primary \(\mathrm{STH}_{j_e}\) to the live head \(\mathrm{STH}_N\) links checkpoints to the current log.
Asymptotically, proof length becomes \(\log_2 |S_e| + \log_2 k\) instead of \(\log_2 N\).

\paragraph{Optional attestation.}
Checkpoints may carry an aggregated BLS signature over a canonical checkpoint digest by agents whose events appear in the covered segments, serving as a compact “stamp of inclusion.” Key management and aggregation mechanics are out of scope and are only one of several viable strategies.

\paragraph{Properties \& trade-offs.}
\begin{enumerate}[itemsep=0pt,parsep=0pt,topsep=0pt,partopsep=0pt]
  \item \textbf{Non-destructive:} the primary log is never pruned or rewritten.
  \item \textbf{Sound \& monotone:} later checkpoints extend coverage without invalidating earlier proofs.
  \item \textbf{Benefit focus:} historical ranges gain shorter proofs; very recent leaves (inside the lag) still use primary proofs.
  \item \textbf{Operational cost:} the secondary index increases PS complexity but requires no event-format changes.
\end{enumerate}
\paragraph{Scope.}
This optimization is proposed for future work and is not essential to the core protocol in this paper.

\section{Conclusion}
We introduced a lightweight model for preserving NHI call-chain integrity in high-traffic multi-agent systems. By combining A2A discovery for identifiers, Merkle proofs for lineage inference, DPoP for cross-org replay resistance, and tamper-evident provenance ledgers for audit, our design preserves verifiable end-to-end trust without bloating tokens. Future work includes integrating purpose-based access control and extending provenance proofs to multi-domain federations.

\clearpage

\appendix

\section{Agent Cards}
\label{appendix:agentcards}

To complement the FedRAMP workflow described in Section~\ref{sec:workflow},this appendix provides the JSON-based \emph{Agent Cards} for each of the autonomous agents (A1–A5). These cards capture the agent’s metadata, capabilities, and identity bindings in a machine-verifiable format.
They are included here as reference artifacts and are not required
to follow the main text.


\lstset{language=json, caption={Readiness Coordinator Agent (A1)}, label={lst:A1}}
\begin{minipage}{\linewidth}
\begin{lstlisting}
{
  "protocolVersion": "0.3.0",
  "name": "readiness-coordinator",
  "description": "FedRAMP readiness coordinator agent that initiates compliance workflows by generating cryptographically signed readiness_start events that reference AO boundary approval, ensuring automation only proceeds under authorized conditions",
  "url": "https://agents.fedramp.gov/readiness-coordinator",
  "preferredTransport": "JSONRPC",
  "version": "1.0.0",
  "capabilities": {
    "streaming": false,
    "pushNotifications": true
  },
  "skills": [
    {
      "id": "initiate-workflow",
      "name": "Workflow Initiation",
      "description": "Initiates FedRAMP compliance workflows based on signed AO boundary approval, serving as the first agent-driven lineage entry"
    },
    {
      "id": "validate-boundary-approval",
      "name": "Boundary Approval Validation",
      "description": "Validates cryptographic signatures on boundary approval documents from Authorizing Officials before proceeding"
    },
    {
      "id": "generate-readiness-event",
      "name": "Readiness Event Generation",
      "description": "Generates cryptographically signed readiness_start events that bind automation sequence to human decisions"
    }
  ],
  "provider": {
    "name": "FedRAMP Authority",
    "domain": "fedramp.gov"
  },
  "identity": {
    "agent_id": "aid://sha256(2A393A61673A7980A1B2C7FCBA0A9C2A957D7397F6C78F22804610A5F4070AD4||fedramp.gov||1758075040)",
    "public_key": "ed25519:DvQgYrpYTgaKB/YDFNoc+ztyYIy7hbxgTB9pmcHkhow=",
    "identity_proof": "ed25519:BC5BE98B05B8488178F7BA78BFEDEF491CC146BEA576DB964D9AF1B8B3...",
    "lineage_support": {
      "merkle_proof_generation": true,
      "dpop_binding": true
    }
  }
}
\end{lstlisting}
\end{minipage}

\lstset{language=json, caption={Evidence Harvester Agent (A2)}, label={lst:A2}}
\begin{minipage}{\linewidth}
\begin{lstlisting}
{
  "protocolVersion": "0.3.0",
  "name": "evidence-harvester",
  "description": "Evidence collection agent that interfaces with system-of-record APIs (AWS Config, Azure Resource Graph, GCP Asset Inventory) to obtain authoritative asset inventories, IAM policies, and configuration baselines, ensuring evidence originates from trusted infrastructure sources",
  "url": "https://agents.fedramp.gov/evidence-harvester",
  "preferredTransport": "JSONRPC",
  "version": "2.1.0",
  "capabilities": {
    "streaming": true,
    "pushNotifications": false
  },
  "skills": [
    {
      "id": "collect-inventory",
      "name": "Asset Inventory Collection",
      "description": "Queries cloud provider APIs to collect comprehensive asset inventories with cryptographic hashing for tamper detection"
    },
    {
      "id": "harvest-iam-policies",
      "name": "IAM Policy Harvesting",
      "description": "Extracts Identity and Access Management policies from cloud environments for compliance analysis"
    },
    {
      "id": "baseline-configuration",
      "name": "Configuration Baseline Collection",
      "description": "Gathers system configuration baselines from authoritative infrastructure sources to establish security posture"
    },
    {
      "id": "hash-artifacts",
      "name": "Artifact Hashing",
      "description": "Generates cryptographic hashes of collected evidence to ensure integrity and anchor into lineage log"
    }
  ],
  "provider": {
    "name": "FedRAMP Authority",
    "domain": "fedramp.gov"
  },
  "identity": {
    "agent_id": "aid://sha256(07372DF22B8528A4B7015B6F72D92B62B9B1C0705E9CB8FE693B17AACF3C4E6E||fedramp.gov||1758075040)",
    "public_key": "ed25519:ycbVVaBqXXomnQSv2U7VieUIdLwrxPGLIjqhjICaEOE=",
    "identity_proof": "ed25519:0B8A5D2D07730EC3B8604872B8CE02802C328D60BC8C427B9181211A63...",
    "lineage_support": {
      "merkle_proof_generation": true,
      "dpop_binding": true
    }
  }
}
\end{lstlisting}
\end{minipage}

\lstset{language=json, caption={Scanner Orchestrator Agent (A3)}, label={lst:A3}}
\begin{minipage}{\linewidth}
\begin{lstlisting}
{
  "protocolVersion": "0.3.0",
  "name": "scanner-orchestrator",
  "description": "Vulnerability and configuration scanning agent that executes systematic scans against inventory defined by Evidence Harvester, producing signed, hashed results for immutable proof that scans were executed against the actual system boundary",
  "url": "https://agents.fedramp.gov/scanner-orchestrator",
  "preferredTransport": "JSONRPC",
  "version": "1.5.2",
  "capabilities": {
    "streaming": true,
    "pushNotifications": true
  },
  "skills": [
    {
      "id": "vulnerability-scan",
      "name": "Vulnerability Scanning",
      "description": "Orchestrates comprehensive vulnerability scans using multiple scanning engines against target inventory"
    },
    {
      "id": "configuration-scan",
      "name": "Configuration Compliance Scanning",
      "description": "Performs configuration compliance checks against FedRAMP baseline requirements and security controls"
    },
    {
      "id": "scan-orchestration",
      "name": "Scan Orchestration",
      "description": "Coordinates multiple scanning tools and engines to ensure systematic and reproducible scan execution"
    },
    {
      "id": "result-signing",
      "name": "Scan Result Signing",
      "description": "Cryptographically signs scan results and produces tamper-evident hashes for lineage verification"
    }
  ],
  "provider": {
    "name": "FedRAMP Authority",
    "domain": "fedramp.gov"
  },
  "identity": {
    "agent_id": "aid://sha256(D779B2B318835CA262B7AC308A90DE37F66611DF8A31692C1B57FBC53A25E82E||fedramp.gov||1758075040)",
    "public_key": "ed25519:rqfSyxjqgDwzrVWy/D62f1l/oaq5qEXIf8wFP+996Nk=",
    "identity_proof": "ed25519:D2D6FEF275E1E3D8625350F49150062615F181CB90BAB59293997874BA...",
    "lineage_support": {
      "merkle_proof_generation": true,
      "dpop_binding": true
    }
  }
}
\end{lstlisting}
\end{minipage}

\lstset{language=json, caption={SSP Packager Agent (A4)}, label={lst:A4}}
\begin{minipage}{\linewidth}
\begin{lstlisting}
{
  "protocolVersion": "0.3.0",
  "name": "ssp-packager",
  "description": "System Security Plan assembly agent that mechanically binds SSP documentation to verifiable, logged evidence from inventory and scan events, preventing drift between documentation and actual system state while providing auditors with cryptographically anchored assurance",
  "url": "https://agents.fedramp.gov/ssp-packager",
  "preferredTransport": "JSONRPC",
  "version": "3.0.1",
  "capabilities": {
    "streaming": false,
    "pushNotifications": false
  },
  "skills": [
    {
      "id": "assemble-ssp",
      "name": "SSP Assembly",
      "description": "Assembles System Security Plan sections by citing and binding to prior evidence events from inventory and scanning"
    },
    {
      "id": "evidence-citation",
      "name": "Evidence Citation",
      "description": "Creates verifiable citations linking SSP content to specific evidence events in the lineage log"
    },
    {
      "id": "document-generation",
      "name": "Document Generation",
      "description": "Generates structured SSP documentation in FedRAMP-compliant formats with embedded evidence references"
    },
    {
      "id": "integrity-binding",
      "name": "Evidence Integrity Binding",
      "description": "Ensures SSP content is cryptographically bound to underlying evidence to prevent documentation drift"
    }
  ],
  "provider": {
    "name": "FedRAMP Authority",
    "domain": "fedramp.gov"
  },
  "identity": {
    "agent_id": "aid://sha256(87549AC5D58A20428E17D074F382C75E39426964B56B2E50289304A31B907271||fedramp.gov||1758075040)",
    "public_key": "ed25519:GxDrnS5fO/XPju5vrTmTomDc/citc+cvzLDHkkiHIKQ=",
    "identity_proof": "ed25519:23F594085F30913AB9DF703B92B7D1FCE00486AF5FA8B6949FB6026095...",
    "lineage_support": {
      "merkle_proof_generation": true,
      "dpop_binding": true
    }
  }
}
\end{lstlisting}
\end{minipage}

\lstset{language=json, caption={3PAO Liaison Agent (A5)}, label={lst:A5}}
\begin{minipage}{\linewidth}
\begin{lstlisting}
{
  "protocolVersion": "0.3.0",
  "name": "3pao-liaison",
  "description": "Third-Party Assessment Organization liaison agent that packages collected evidence (inventory, scans, SSP) with Merkle inclusion proofs into auditable evidence capsules for delivery to independent assessors, providing verifiable provenance without altering evidence",
  "url": "https://agents.fedramp.gov/3pao-liaison",
  "preferredTransport": "JSONRPC",
  "version": "1.2.3",
  "capabilities": {
    "streaming": false,
    "pushNotifications": true
  },
  "skills": [
    {
      "id": "package-evidence",
      "name": "Evidence Capsule Packaging",
      "description": "Packages evidence artifacts with their corresponding Merkle inclusion proofs into tamper-evident capsules"
    },
    {
      "id": "generate-proofs",
      "name": "Inclusion Proof Generation",
      "description": "Generates Merkle inclusion proofs demonstrating evidence integrity against transparency log"
    },
    {
      "id": "capsule-delivery",
      "name": "Secure Capsule Delivery",
      "description": "Delivers evidence capsules to external assessors with cryptographic integrity guarantees"
    },
    {
      "id": "provenance-tracking",
      "name": "Provenance Tracking",
      "description": "Maintains verifiable chain of custody from evidence collection through external assessment handoff"
    }
  ],
  "provider": {
    "name": "FedRAMP Authority",
    "domain": "fedramp.gov"
  },
  "identity": {
    "agent_id": "aid://sha256(4A957DF836DDCD0E46F31625F354E0B0ABB24304E3B33B21E779ACD375C8B87A||fedramp.gov||1758075040)",
    "public_key": "ed25519:IdIy83UIRJR09Nun/MWx91ct6vYoErvJRXX+uw4RvRU=",
    "identity_proof": "ed25519:681977929BAE341608CBAEE6C0E6EF80BD2E8C5A69A07CF568BEF2160...",
    "lineage_support": {
      "merkle_proof_generation": true,
      "dpop_binding": true
    }
  }
}
\end{lstlisting}
\end{minipage}

\end{document}